\documentclass[aps,prd,twocolumn,showpacs,groupedaddress]{revtex4-1}
 
\pdfoutput=1 
\usepackage{amssymb}
\usepackage{amsmath}
\usepackage{hyperref}
\usepackage{graphicx}
\usepackage{float} 

\begin{document}

\title{First law and a variational principle for \\ static asymptotically Randall-Sundrum black holes}

 \author{Scott Fraser}
\email[Email address: ]{scfraser@calpoly.edu}
\affiliation{Department of Physics, California Polytechnic State University, San Luis Obispo, California 93407, USA}

\author{Douglas M. Eardley}
\email[Email address: ]{doug@kitp.ucsb.edu}
\affiliation{Department of Physics, University of California, Santa Barbara, California 93106, USA}

\begin{abstract}
We   give a new, intrinsic,  mass definition for spacetimes asymptotic to    the Randall-Sundrum  braneworld models, RS1 and RS2.  For this mass, we  prove a      first law  for static black holes,  including
variations of the  bulk  cosmological constant, 
brane tensions, and    RS1 interbrane distance.  
Our first law defines  a thermodynamic volume
and  a gravitational tension that are braneworld analogs of  the  corresponding quantities in  asymptotically AdS
  black hole spacetimes   and asymptotically flat
  compactifications,  respectively. 
We also  prove
the following related variational principle for asymptotically RS
black holes:
 instantaneously static initial data that extremizes
the mass yields a static black hole,
for variations  at 
 fixed     apparent horizon area, AdS
curvature length, cosmological constant,  brane tensions, and  RS brane warp factors.  
This variational principle is valid  with either  two branes (RS1) or  one brane (RS2), and is  applicable to  variational  trial       solutions.
\end{abstract}

\pacs{04.50.Gh, 04.70.Bw}
 
 % 04.20.Fy deleted
 
\maketitle

\section{Introduction }

Static and stationary black holes should obey  four   classical laws \cite{Bardeen-four-laws}.
The  first law   expresses  conservation of energy, and has been proven     
in  spacetimes with four dimensions  \cite{Bardeen-four-laws, sudarsky-wald} 
and  higher dimensions 
\cite{Gibbons-first-law, bh-enthalpy, Gibbons-enthalpy},
including   a compact extra dimension
 \cite{harmark-obers-def-tension, *harmark-obers-first-law, *kol-sorkin-piran},
 but   not  in
  spacetimes asymptotic to the Randall-Sundrum (RS)  braneworld models \cite{RS1, RS2}.
In  this paper, we   close this gap by proving a  general first law for static asymptotically  RS  black  holes.
The  first law    relates the variations of mass and
 other physical quantities.  
For this purpose, we provide a new,  intrinsic, mass definition.
Our first law       defines  
a thermodynamic volume
and  a gravitational tension  that are   braneworld analogs of   
thermodynamic volume  in 
  asymptotically AdS
  black hole spacetimes \cite{bh-enthalpy, Gibbons-enthalpy}
and gravitational tension  in
asymptotically flat
  compactifications
 \cite{harmark-obers-def-tension, *harmark-obers-first-law, *kol-sorkin-piran}.

The RS models   are phenomenologically interesting, and  have holographic interpretations \cite{RS-AdS-CFT} in   the AdS/CFT  correspondence.
In the RS models, our observed universe is
a brane surrounded by an AdS bulk. The bulk is warped by  a negative  
cosmological constant.
The RS1 model \cite{RS1}  has two branes of opposite tension, with
our universe  on the negative-tension brane.  
 Tuning the interbrane
distance   appropriately predicts the 
   production of small black holes at  TeV-scale collider energies 
    \cite{Banks-Fischler, *Giddings}, and   LHC experiments  
  \cite{LHC-RS-extra-dimensions, *LHC-RS-black-holes} continue to test this hypothesis. 
 In the RS2 model \cite{RS2}, 
our  universe resides on  the positive-tension brane,
with the negative-tension brane  removed to infinite distance.
Perturbations of RS2   reproduce
 Newtonian gravity  at large distance  on the brane, while in
 RS1 this requires a mechanism to stabilize the interbrane distance \cite{Tanaka}.
In  RS2,   solutions for static black holes    on the brane   have been found  numerically,   for both
    small black holes
\cite{Kudoh-smallBH-1, *Kudoh-smallBH-2, *Kudoh-smallBH-6D} and large  
 black holes  \cite{Figueras-Wiseman, *Abdolrahimi}, compared to the AdS curvature length.
The only known exact  analytic black hole
solutions
are the static and stationary  solutions   \cite{ehm, *ehm-2} in a lower-dimensional version of RS2.
An exact solution for a large black hole on the brane was also found in \cite{nonempty-bulk} in a generalized RS2 setup with matter in the bulk.

In general, 
the first law   for a static or stationary black hole  
 takes the    form
 $\delta M = (\kappa/8\pi G) \delta A + \sum_i p_i \,\delta Q_i$.  
This relates the variations of mass $M$, horizon area $A$,
and other physical quantities $Q_i$. Thus, if a  black hole is  static or stationary,    it   
extremizes  $M$  under variations  that hold  constant the
remaining variables  ($A$, $Q_i$). 
The  converse        of this statement
motivates  a   variational principle:
{\it If a black hole's exterior spatial geometry  is  initially   static (or    initially stationary) and
extremizes the mass  with other physical variables held fixed, then  
the black hole
is static  (or stationary).} 
In this variational principle, the  specific  variables  to   hold fixed  depend on    
the  form of the  first law. The  appropriate area to hold fixed is that of the 
black hole  apparent horizon, which 
  is   determined   by the  
  spatial geometry alone (unlike   the   event horizon, which is a global  spacetime property).
The apparent horizon
 generally lies inside  the event horizon, and  coincides with it
  for a static or stationary black hole spacetime.

For asymptotically flat black holes in four spacetime dimensions, 
 a version of the above   variational principle    was proved
 by Hawking for stationary black holes \cite{hawking}, and was extended to Einstein-Yang-Mills theory 
  by Sudarsky and Wald   
 \cite{sudarsky-wald, chrusciel-wald}.  In this paper, we prove   a version of the above
 variational principle for static asymptotically RS black holes. The quantities held fixed in our variational principle are 
   the AdS curvature length, cosmological constant, brane tensions, 
and RS values (at spatial infinity) of   warp factors   on each brane.
The variations of these quantities   appear in the general
  first law  that we   prove in this paper.

This   paper is organized as follows.  After reviewing the RS spacetimes in section \ref{RS-review}, 
we   define the mass  for an asymptotically RS spacetime in section \ref{sec:mass}, and evaluate the mass
for a static asymptotic solution in section \ref{sec:static-soln}.   We prove  the first law for  static black holes in
section \ref{sec:first-law}.  
We  prove the variational principle in section \ref{vp-proof}, including
an explicit  application using a trial solution.  We conclude in section \ref{sec:conclusion}.

\begin{figure}[H]
\centering
\includegraphics{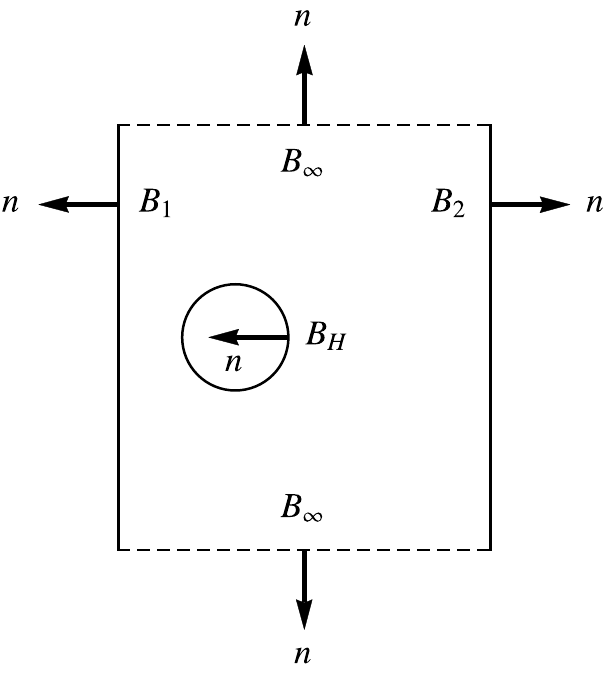}
\caption{\label{initial-data-figure}   Illustration of a spatial hypersurface $\Sigma$, for a  black hole with apparent horizon  $B_H$  not intersecting the branes,  $B_1$ and $B_2$.  Spatial infinity $B_\infty$ is a single boundary, transverse to both branes.  Each boundary $B$ has outward normal $n$.
   } 
\end{figure}

Throughout this paper, we   use two branes, so  our results
 apply to 
     either   RS1   or   RS2  
in the appropriate limit.
We work on the orbifold  region  (between the branes) and  use  $D$ spacetime dimensions. 
A timelike surface has metric $\gamma_{ab}$, extrinsic curvature
$K_{ab}=\gamma_a{}^c\nabla_c n_b$, and outward unit normal     $n_b$.
A spatial hypersurface   $\Sigma$
has unit normal $u_a$, metric $h_{ab}$, and covariant derivative $D_a$.
Each  boundary $B$ of $\Sigma$
  has metric  $\sigma_{ab}$, extrinsic curvature $k_{ab}=h_a{}^c D_c n_b$,
   and outward unit normal     $n_b$.
The  boundaries $B$ of $\Sigma$  are illustrated in Fig.\ \ref{initial-data-figure}.
These  boundaries  are:
 spatial infinity $B_\infty$,
 the branes ($B_1$, $B_2$), 
 and 
 the black hole apparent horizon  $B_{H}$.  If the  black hole is static, then
   $B_{H}$
  coincides with the black hole event horizon in $\Sigma$.

\section{The Randall-Sundrum spacetimes
\label{RS-review}}

The RS spacetimes  \cite{RS1,RS2} are portions of 
 an anti-de Sitter  (AdS) spacetime, with metric
\begin{equation}
\label{g_RS}
ds^2_{\rm RS}
= 
\Omega(Z)^2 \left(- dt^2 + d\rho^2 + \rho^2d\omega_{D-3}^2 + dZ^2 \right) \ .
\end{equation}
Here $d\omega_{D-3}^2$ denotes the unit $(D-3)$-sphere.  
The    warp factor is   $\Omega(Z)=\ell/Z$,  with values   $\Omega_i$ on each brane.
Here $\ell$ is the AdS curvature length, related to the bulk cosmological constant $\Lambda<0$    given below.
The RS1 model \cite{RS1} contains two  branes, which are the    surfaces      $Z=Z_i$ 
with brane  tensions  $\lambda_i$, where $i=1,2$. 
The brane tensions $\lambda_i$ and      bulk cosmological constant  $\Lambda$ are  
  \begin{equation}
  \label{RS-tensions}
\lambda_1 =  -\lambda_2    =
\frac{2(D-2)}{8\pi G_D\ell}
\ , \quad
\Lambda    =   -\,\frac{(D-1)(D-2)}{2\ell^2} \ .
\end{equation} 
The dimension $Z$ is   compactified on the orbifold $S^1/\mathbb{Z}_2$ and the 
  branes have    orbifold mirror
 symmetry:
in the covering space,  symmetric points across a brane are identified.  
There is a  discontinuity   in the extrinsic curvature $K_{ab}$ across each brane
 given by the Israel condition  \cite{israel}. Using
 orbifold symmetry, the Israel condition
 requires  the   extrinsic curvature at each brane to satisfy
 \begin{equation}
\label{RS-K}
2 K_{ab}
 = \left(\frac{8\pi G_D\lambda}{D-2}\right) \gamma_{ab}
\ , \quad
2k_{ab} = \left(\frac{8\pi G_D\lambda}{D-2}\right)  \sigma_{ab} \ .
\end{equation}
Using (\ref{RS-tensions}), this can also be written as
\begin{equation}
2 K_{ab}
 = \frac{\varepsilon}{\ell}\gamma_{ab}
\ , \quad
2k_{ab} = \frac{\varepsilon}{\ell}  \sigma_{ab} \ ,
\end{equation}
where   $\varepsilon=\pm 1$  is   the sign of each brane tension.
The RS2 spacetime \cite{RS2}
is   obtained   from RS1
 by   removing the  negative-tension brane (now   a regulator)
 to infinite distance ($Z_2 \rightarrow \infty$)
and the orbifold  region has $Z \ge Z_1$.

\section{Mass definition
\label{sec:mass}}

  For an asymptotically RS spacetime, we   will  define the mass $M$  
   using a counterterm method
\cite{balasubramanian-kraus, *kraus-larsen-siebelink}.
This  is an intrinsic approach,  which is
well suited to the
  variations   we   will perform  in section \ref{sec:first-law} to prove the first law, and
   is also useful for spacetimes with nontrivial topology.   
By comparison, other  definitions, such as  the
   Brown-York mass \cite{brown-york}, use an auxiliary  
reference   spacetime, which may be
awkward when the topology is nontrivial.  
A reference spacetime  approach is also implicit 
when the asymptotic geometry
serves this purpose, as in the
  Abbott-Deser  mass    \cite{abbott-deser} and its counterpart
  with an asymptotically flat  compact dimension, the
  Deser-Soldate  mass \cite{Deser-Soldate}. 
 For an asymptotically RS spacetime,
it is straightforward to verify that evaluating our mass definition,  as in   (\ref{mass}) below,
reduces to  the same result as the
Abbott-Deser mass  formula \cite{abbott-deser}.

In the counterterm approach \cite{balasubramanian-kraus, *kraus-larsen-siebelink},   for a spacetime   with metric $g_{ab}$,
one first evaluates   the bare action $\widetilde S$.
If  this diverges,  one constructs
an action counterterm $S_{ct}$
to render the sum  $S=\widetilde S+S_{ct}$ finite, as  follows.
Let the metric $g_{ab}$
asymptote to   $g_{ab}^{(0)}$  whose bare
action
$\widetilde S_0$ also diverges.
We express $\widetilde S_0$
in terms of its intrinsic boundary invariants, and
 define the action counterterm
$S_{ct}=-\widetilde S$ where $\widetilde S$ is
the same functional of its boundary geometry
 that
$\widetilde S_0$ is of its boundary geometry.
This gives   $S_0=0$ for  $g_{ab}^{(0)}$  and      a finite action  
  $S$ for   $g_{ab}$.
  
To define the mass $M$, one     proceeds from the action to the Hamiltonian
  (defined on an arbitrary initial value
 spatial hypersurface $\Sigma$),
which is given by a bulk term involving
 initial value constraints, and
surface terms.
For a solution to the constraints,
the bulk term vanishes  and the
bare mass at spatial infinity is 
 \cite{brown-york}
\begin{equation}
\label{M-bare-formula}
\widetilde M =    - \, \frac{1}{8\pi G_D}\int_{B_\infty} d^{D-2}x\,N
\sqrt{\sigma} \, k \ .
\end{equation}
Here  $k$ is the extrinsic curvature of the boundary
$B_\infty$ and for a static spacetime,  the lapse function is $N=\sqrt{-g_{tt}}$.
If $\widetilde M$  diverges,     one   constructs a mass
counterterm
\begin{equation}
\label{M-ct-formula}
M_{ct} = \int_{B_\infty} d^{D-2} x \, N\sqrt{\sigma} \, u_a u_b \left[
\frac{2}{\sqrt{-\gamma}} \frac{\delta
S_{ct}}{\delta\gamma_{ab}} \right]
\end{equation}
 such that the mass $M$ is finite, defined by
\begin{equation}
\label{finite M}
M = \widetilde M + M_{ct} \ .
\end{equation}
Following the above procedure, we   begin with the RS spacetime.
The bare action $\widetilde S_{\rm RS}$ consists of a bulk term $S_\Sigma$ and a Gibbons-Hawking term
at each boundary,
\begin{equation}
\label{S_RS}
   \widetilde S_{\rm RS} = S_{\Sigma}   + S_1  +  S_2  +  S_\infty  \ .
\end{equation}
Here
\begin{eqnarray}
\nonumber
S_{\Sigma} &=& \frac{1}{16\pi G_D} \int dt \int_{\Sigma}
d^{D-1}x\,\sqrt{-g}\left( R -2\Lambda\right) \ ,
\\
\nonumber
S_i &=&  \frac{1}{8\pi G_D}\int dt \int_{B_i} d^{D-2}x\,\sqrt{-\gamma}
\left(K-\frac{8\pi G_D\lambda_i}{2}\right) \ ,
\\
S_\infty &=&  \frac{1}{8\pi G_D} \int dt  \int_{B_\infty} d^{D-2}x\,\sqrt{-\gamma}\,
K    \ .
\end{eqnarray}
For the RS solution  (\ref{g_RS}),  the   Ricci scalar is   $R=2\Lambda D/(D-2)$ 
and  after integrating  $ S_{\Sigma}$ in $Z$, we find 
   \begin{equation}
 S_{\Sigma} + S_1 + S_2 = 0 \ .
\end{equation}
   We now specialize to the case $D=5$, for which
   \begin{equation}
 \label{S-RS-1} \widetilde S_{\rm RS} 
 = S_\infty 
 = \frac{1}{8\pi G_5}  \int d^{4}x\,\sqrt{-\gamma}\, \frac{2}{\Omega \rho} 
\ .
\end{equation}
This   diverges as
 $\rho \rightarrow \infty$. 
   The metric $\gamma_{ab}$ on this boundary   is  (\ref{g_RS}) with $\rho=$ constant.
Let $\hat\gamma_{ab}$ be the submetric with $Z=$ constant, 
and  let
 $\hat\sigma_{ab}$  be  the submetric on a 2-sphere (constant $\rho$, $Z$, $t$).
Their Ricci scalars are 
   \begin{equation}
\hat{\cal R}(\hat\gamma)  =\hat{\cal R}(\hat\sigma) 
  =  \frac{2}{(\Omega \rho)^2 }\ .
\end{equation}
If we express   $\widetilde S_{\rm RS}$
 in terms of    $\hat{\cal R}$, then  
 an   asymptotically RS
spacetime  has action
counterterm  
$
 S_{ct}  =  -\widetilde S
$
 where $\widetilde S$ is
the same functional of its boundary geometry
that
$\widetilde S_{\rm RS}$ is of its   geometry.  Thus
\begin{equation}
\label{S-ct-hybrid}
S_{ct}  =  - \, \frac{\sqrt{2}}{8\pi G_5} \int d^4x\,
\sqrt{-\gamma}\,\sqrt{\hat{\cal R}}
  \ .
\end{equation}
All   quantities in (\ref{S-ct-hybrid})
  refer to the boundary geometry of a general metric  $g_{ab}$,
  not the RS metric  (\ref{g_RS}).  
As shown in Appendix \ref{appendix-M-ct},   the mass counterterm   (\ref{M-ct-formula}) then
 yields our  mass definition     
for an asymptotically RS spacetime, 
\begin{equation}
\label{mass-geometric}
M =
\frac{1}{8\pi G_5}
 \int_{B_\infty} d^{3} x\, N \sqrt{\sigma} \left(- k + \sqrt{2\hat{\cal R}}\right)
\ .
\end{equation}

\section{Static asymptotic solution
\label{sec:static-soln}}

Here we   evaluate our mass definition   (\ref{mass-geometric})
for a static asymptotic solution, which we will use 
 in section \ref{sec:first-law} below.
We define the  branes as the surfaces $Z=Z_1$ and $Z=Z_2$.
We consider  a
 static     asymptotically RS metric   with 
  functions $F_\nu$  that fall off   at large $\rho$ as follows,
\begin{eqnarray}
 \nonumber
 ds^2 &=&
\Omega^2 \left(-e^{2F_t}dt^2 +
e^{2F_\rho}d\rho^2 +   e^{2F_\omega}{\rho^2d\omega_{2}^2}  +  e^{2F_Z}dZ^2 \right) 
 \\
 \label{ansatz-1}
F_\nu &=& \frac{a_\nu(Z)}{\rho} +
\frac{b_\nu(Z)}{\rho^2} + \frac{c_\nu(Z)}{\rho^3}   
 + O(1/\rho^4)\ .
\end{eqnarray}
For these asymptotics, the  mass  (\ref{mass-geometric}) evaluates to
\begin{equation}
  M =
\label{mass}
\frac{1}{G_5} \int_{Z_1}^{Z_2} dZ\, \Omega^3
\left(  a_\rho +    \frac{a_Z}{2}
\right)   \ .
\end{equation}
The value of  $M$ also
appears  in  the  solution to (\ref{ansatz-1}).
To see this, 
we find it is necessary  to solve the Einstein   equations
through third order.
The first order solutions   $a_\nu$  
are
 \begin{subequations}
\label{Order1-3}
    \begin{eqnarray}
\label{a_t}
 a_t(Z) &=& a_\rho(Z)  + \mu_0 \ ,
\\
 a_\omega(Z) &=& a_\rho(Z) + \mu_1 \ ,
\\
 a_Z(Z) &=& -\,Z \,a_\rho^\prime  \ .
    \end{eqnarray}
 \end{subequations}
Here $\mu_0$ and $\mu_1$  are integration
constants and $^\prime=d/dZ$. The constant $\mu_1$  can be removed
by a gauge transformation $\rho \rightarrow \rho + \mu_1/2$.
Two identities we will need are
\begin{subequations}
\begin{align}
\label{identity1}
\int_{Z_1}^{Z_2} dZ\,\Omega^3\left(a_t + a_\rho + a_Z \right) &=  0 \ ,
\\
\label{int-field} \int_{Z_1}^{Z_2} dZ\,\Omega^{n+1} \left(n \,a_\rho +
 a_Z \right)
&= 
- \ell
\left(\Omega^{n}  a_\rho\right)\Big|_{Z_1}^{Z_2} \ .
\end{align}
\end{subequations}
The identity (\ref{int-field}) also holds with $a_\rho$   replaced by $a_t$ or
$a_\omega$. Using  (\ref{int-field}), the mass  can be written as
\begin{equation}
 \label{M-local}
 M =  -\,\frac{\ell}{2G_5}\left(\Omega^2 a_\rho\right)\Big|^{Z_2}_{Z_1}
 \ .
\end{equation}
The third order
  solutions $c_\nu$   involve an
integration constant $q_0$.  
The Israel condition  provide one equation at each brane, which can be solved for  $\mu_0$ and  $q_0$ as
\begin{eqnarray}
\label{mu_0}
\mu_0\left(\Omega_2{}^2-\Omega_1{}^2\right)
 &=& 
-2\left(\Omega^2 a_\rho\right)\Big|^{Z_2}_{Z_1} \ ,
\\
\label{q_0}
 q_0 (\Omega_2^{-2} -\Omega_1^{-2})  &=& 
 2 a_\rho\Big|_{Z_1}^{Z_2}
 \ .
\end{eqnarray}
Here $\Omega_i=\ell/Z_i$   the warp factor at each brane.
We see from  (\ref{int-field})  and (\ref{mu_0})
that   $M$  is proportional to    $\mu_0$, 
  \begin{equation}
  \label{M and mu0}
 M =  \frac{\ell\mu_0}{4G_5} \left(\Omega_2{}^2-\Omega_1{}^2\right)
 \ .
\end{equation}
We see  from  (\ref{q_0}) that   $q_0$ is proportional to a   quantity
 $\mathcal{Q}$ 
   parametrizing the      interbrane distance    near $\rho \rightarrow \infty$,
\begin{equation}
{\cal L}_{\rm branes}
= L   +         \frac{\mathcal{Q}}{\rho}   +   O(1/\rho^2)  \ ,
\end{equation}
where the distance $L$ at infinity  and the  constant $\cal Q$ are
\begin{equation}
\label{L asymp}
L  
 =  \ell\ln \left(\frac{\Omega_1}{\Omega_2}\right)
\quad , \quad
\mathcal{Q} = -\ell a_\rho\Big|^{Z_2}_{Z_1} \ .
\end{equation}
We will refer to $L$ and $\cal Q$ in the next section.
We will also use the fact that
  $M$ and $\cal Q$ appear in the values of
 $a_\rho$  at each brane,   which we  find
 by  solving
 (\ref{M-local}) and (\ref{L asymp}),  
 \begin{equation}
   \label{a-M-Q}
 a_\rho(Z_i)=\frac{2G_5M -\Omega_j{}^2 {\cal Q}}{\ell(\Omega_1{}^2-\Omega_2{}^2)}
\quad , \quad j \neq i  \ .
 \end{equation}
In the RS1 case,     $M$ and $\cal Q$    
   have    lower-dimensional
interpretations on each brane.  This follows since
there is an effective Brans-Dicke  gravity     on  each brane \cite{garriga-tanaka},
and 
Brans-Dicke gravity contains  two  asymptotic quantities, a
 tensor mass and  a scalar mass \cite{Lee-BD-masses}.  On each brane, one can verify that  
 $M$ and $\cal Q$
 are proportional to the  effective tensor  mass and    scalar mass, respectively.

\section{First law for    static black holes
\label{sec:first-law}}

\subsection{Preliminary form}

For a static or stationary black hole, 
the  first law   relates the   variations  of  mass, black hole horizon area, and other physical quantities. 
We will 
 include
variations of the  bulk  cosmological constant $\Lambda$ and
brane tensions $\lambda_i$ that preserve the RS  conditions (\ref{RS-tensions}),
\begin{equation}
\label{d-ell}
-\, \frac{\delta\ell}{\ell}
= \frac{\delta\lambda_1}{\lambda_1}
= \frac{\delta\lambda_2}{\lambda_2}
= \frac{\delta\Lambda}{2\Lambda}   \ .
\end{equation}
We will also include the  variation  $\delta L$ of the   interbrane separation.  From (\ref{L asymp}),
\begin{equation}
\label{dL} 
\delta L = \delta\ell\,\ln\left(\frac{\Omega_1}{\Omega_2}\right)
+ \ell \, \frac{\delta\Omega_1}{\Omega_1} 
-\ell \, \frac{\delta\Omega_2}{\Omega_2}
\ .
\end{equation}

  Our setup is general: it applies to a black hole  localized on   a brane,   or    isolated    in the bulk (away from   either brane),  and also applies to the asymptotically RS black string \cite{black-string}.
Our method    is based on the Hamiltonian approach of 
  \cite{sudarsky-wald},    with  the additional
  variations (\ref{d-ell})\textendash(\ref{dL}). 
  The full
Hamiltonian contains a bulk term
and boundary terms.  The bulk  term $H_{\Sigma}$ is  defined on
a spatial hypersurface $\Sigma$,
\begin{equation}
\label{H_sigma}
 H_{\Sigma} =\int_{\Sigma} d^{D-1} x\, \left( N
{\cal C}_0 + N^a {\cal C}_{a} \right) \ .
\end{equation}
Here $N$ and $N^a$ are
the   lapse and shift in the standard decomposition of the   spacetime metric.  
 Our focus is the initial data ($h_{ab}$, $p^{ab}$) on $\Sigma$, where  $h_{ab}$  is the spatial metric 
 and
  $p^{ab}$ is its  canonically conjugate momentum,
\begin{equation}
   16\pi G_D \, p^{ab}=\sqrt{h} {\cal K}^{ab} - {\cal K}\, h^{ab} 
   \quad , \quad
    {\cal K}_{ab} = h_a{}^c \nabla_c u_b
    \ .
\end{equation}
Initial data must satisfy     constraints, 
${\cal C}_0 = 0$ and ${\cal C}_a= 0$, which we henceforth assume, where
\begin{eqnarray}
 \nonumber {\cal C}_0 &=& 
\frac{\sqrt{h}}{16\pi G_D}\left(2\Lambda - {\cal R}\right)  +
\frac{16\pi G_D}{\sqrt{h}} \left( p^{ab} p_{ab} - \frac{ p^2}{D-2}
\right) \ ,
\\
 {\cal C}_a &=&  -2 D_b p_a{}^b \ .
 \label{constraints}
\end{eqnarray}
Here
$\cal R$ and $D_a$ are the Ricci scalar and
  covariant derivative
associated 
with $h_{ab}$.
We now consider the change  $\delta  H_{\Sigma}$ under
 variations  ($\delta h_{ab}$, $\delta p^{ab}$).  
One finds  
 $\delta{\cal C}_0$ and $\delta{\cal C}_a$
involve derivatives ($D_c\delta h_{ab}$, $D_c\delta p^{ab}$).
Integrating  by parts to remove these
derivatives yields
  surface terms  $I_B$,  
\begin{eqnarray}
\nonumber
  \delta H_{\Sigma} &=&    \int_{\Sigma} d^{D-1}x
 \left[
 {\cal P}^{ab}\delta h_{ab} + {\cal H}_{ab}\, \delta p^{ab}
 \right]
\\
& &
+ \
\frac{\delta\Lambda}{8\pi G_D}
\int_{\Sigma} d^{D-1}x \,
N \sqrt{h}
\ + \
\sum_B I_B
\label{delta_H_sigma-1}
\ .
\end{eqnarray}
 The sum in (\ref{delta_H_sigma-1}) is over  all of the boundaries $B$ illustrated in Fig.\ \ref{initial-data-figure}.
The quantities  ${\cal P}_{ab}$ and ${\cal H}_{ab}$
appear in the  time  evolution equations,
 \begin{equation}
 \label{evolution}
 \dot h_{ab} = {\cal H}_{ab} \ , \quad
 \dot p^{ab}=-{\cal P}^{ab} \ ,
\end{equation}
where the overdot
denotes the Lie derivative
along the time evolution vector field $t^a= N u^a + N^a$
with
$u^a$   the unit normal to $\Sigma$.
We will not need the 
most general forms of  ${\cal P}_{ab}$,   ${\cal H}_{ab}$, and  $I_B$.
We will give their  simplified forms below, after implementing some of our key assumptions.

We now assume our variations take
  one solution of the constraints 
 to another solution of the constraints, so we take
  $\delta{\cal C}_0=0$ and $\delta{\cal C}_a=0$.
Then the variation of (\ref{H_sigma})
immediately
  gives     
\begin{equation}
\label{delta_H_sigma-2}
\delta  H_{\Sigma}=0 \ .
\end{equation}
We henceforth assume the initial data  is instantaneously static,  for which
  $p^{ab}=\delta p^{ab}=0$
  and we take $N^a=0$.  
  One then explicitly finds
 ${\cal H}_{ab}=0$, so   (\ref{delta_H_sigma-1}) and   (\ref{delta_H_sigma-2}) give
 \begin{equation}
0  =  \int_{\Sigma} d^{D-1}x
 \left[
 {\cal P}^{ab}\delta h_{ab} 
+ 
\frac{\delta\Lambda}{8\pi G_D}
N \sqrt{h}
\right]
\ + \
\sum_B I_B
\label{delta_H_sigma-1b}
\ .
\end{equation}
This result will be the primary equation for proving our variational principle in section  \ref{vp-proof}.
Here ${\cal P}^{ab}$ is given by
\begin{equation}
 {\cal P}^{ab}   =   \frac{\sqrt{h}}{16\pi G_D}
  \left(
{\cal R}^{ab}+ h^{ab}D_c D^c  - D^a D^b
  \right)N \ .
\end{equation}
We now assume    a static black
hole   with    timelike Killing field $\xi^a$
and  choose  $t^a=\xi^a$.  For a static solution,
  ${\cal P}^{ab}=0$ by (\ref{evolution}).
Then
(\ref{delta_H_sigma-1b})  gives
 \begin{equation}
\label{first-law-1}
 0 =  \frac{\delta\Lambda}{8\pi G_D}\int_{\Sigma} d^{D-1} x\, N
\sqrt{h} \ + \ \sum_B I_B \ .
\end{equation}
This equation is our preliminary form of the first law.  It
 simply  remains    to express (\ref{first-law-1}) in terms of
physical quantities.  
Each  surface term $I_B$  can be written
\cite{brown-lau-york}
\begin{equation}
 I_B =
 \int_{B} d^{D-2}x\, N
 \left[
 \frac{\delta\left(\sqrt{\sigma} k\right)}{8\pi G_D}
   - \frac{\sqrt{\sigma}}{2} s^{ab}\delta\sigma_{ab}
\right]
\label{surface-terms-2}
    -    J_B  \ ,
\end{equation}
where
\begin{align}
 \label{stress}
8\pi G_D\, s^{ab} &=    - k^{ab} + \left[k + n^c
(D_cN)/N\right]\sigma^{ab} \ ,
\\
16\pi G_D J_B &=
  \sum_{B^\prime \not= B}\int_{B \cap B^\prime}
d^{D-3}x \, N\sqrt{\hat\sigma}\,  n^\prime_a\,  \sigma^a{}_b\,
\delta n^b  \ .
\end{align}
Here $\hat\sigma$ denotes  the
metric on $B \cap B^\prime$.  In what follows,   we will have  $J_B=0$.  This 
is  due to
$N=0$ on 
  $B_H$,  and   
 due to  orthogonality ($n^\prime_a n^a=0$) at the other boundaries $B$.

We now evaluate the boundary terms  (\ref{surface-terms-2}) for   $D=5$.
The results at the horizon $B_H$ and the branes $B_i$ are 
\begin{equation}
 \label{da} 
I_{B_{H}} 
=\frac{\kappa}{8\pi G_5}\, \delta A 
 \quad , \quad 
I_{B_i} =
\frac{\delta\lambda_i}{2} \int_{B_i} d^{3}x\,N \sqrt{\sigma} 
\ ,
\end{equation}
with $A$    the black hole horizon area.
These results are straightforward to derive.  
At the black hole horizon $B_H$,  we have $N=0$
and $D_aN = -\kappa\, n_a$ where $\kappa$ is the constant surface gravity \cite{sudarsky-wald}.
This gives
  $8\pi G_5  N s^{ab} = -\kappa\,\sigma^{ab}$ and  the result in (\ref{da}) follows.
 At each brane $B_i$, we
use $n^c (D_cN)/N =K-k$,
which is a  general result  \cite{brown-york}
  valid when $u^a n_a=0$.
Using (\ref{RS-K})
 gives $s^{ab} =(\lambda_i/2)\sigma^{ab}$ which yields  the result  in  (\ref{da}).
In Appendix \ref{appendix-first-law},  we show the boundary term  $I_{B_\infty}$    is
 \begin{equation}
\label{text-I-infty-3}
 I_{B_\infty}  = -  \delta M   
+   {\cal F}_\infty \, \delta \ell
+ {\cal U}_1 \frac{\delta\Omega_1}{\Omega_1} 
- {\cal U}_2 \frac{\delta\Omega_2}{\Omega_2} 
 \ ,
\end{equation}
  where the boundary quantity ${\cal F}_\infty$ at infinity is
   \begin{equation}
    \label{F-infty}
  {\cal F}_\infty =
- \frac{1}{2G_5\ell}\int_{Z_1}^{Z_2} dZ\, \Omega^3\left( a_t  + 2a_Z \right) \ ,
\end{equation}
and the  coefficients $ {\cal U}_i$ are
  \begin{equation}
  \label{U-def}
 {\cal U}_i = 
M\left( \frac{\Omega_i{}^2 }{\Omega_1{}^2-\Omega_2{}^2}\right)
-
\frac{3 {\cal Q}}{2G_5}
\left(
 \frac{\Omega_1{}^2\Omega_2{}^2}{\Omega_1{}^{2}-\Omega_2{}^{2}} 
\right) \ .
\end{equation}
Here  $\cal Q$   parametrizes  the asymptotic      interbrane separation  (\ref{L asymp}). 
We also define $\cal F$ by the following sum,
 \begin{equation}
 \label{F-def}
{\cal F} \,\delta \ell   =
 \frac{\delta\Lambda}{8\pi G_5}\int_{\Sigma} d^{4}x\, N
\sqrt{h}
 +  I_{B_1} +\ I_{B_2}    +    {\cal F}_\infty \delta\ell \ .
\end{equation}
Here the two brane  terms $I_{B_i}$  render the    volume integral  finite, as one can   verify.
The    term ${\cal F}_\infty$  renders
 ${\cal F}$ gauge invariant, as shown in Appendix \ref{gauge invariance}.
We also define ${\cal V}$ by
\begin{equation}
\label{V-def}
 {\cal V} 
 =  \left(\frac{\ell}{2P}\right) {\cal F}
 \quad , 
 \quad
  -\,{\cal V} \, \delta P = {\cal F} \,\delta \ell   \ ,
\end{equation}
where $P = -\Lambda/(8\pi G_5)$ is the pressure due to the cosmological constant.
 We   have now evaluated all the terms needed to  rewrite the preliminary  
  first law
  (\ref{first-law-1}).  

\subsection{The first law}

We will  give four versions of the first law, corresponding to different choices of variations.
Substituting       (\ref{da}),
 (\ref{text-I-infty-3}), and (\ref{V-def}) into   
  (\ref{first-law-1})   gives the first law in the form
 \begin{equation}
   \label{first-law-v1}
   \delta M = \frac{\kappa \,\delta A}{8\pi G_5}
- {\cal V}\, \delta P
+  {\cal U}_1 \frac{\delta\Omega_1}{\Omega_1} 
- {\cal U}_2 \frac{\delta\Omega_2}{\Omega_2} 
\ .
\end{equation}    
The  area  term  is standard.
The last two terms    are changes  in   
mass   due to     changes in the branes' gravitational field, since  $\delta\Omega_i$ are         variations of   gravitational redshift factors on each brane. 
The last term   is absent in RS2,   which removes the negative-tension 
brane   to $Z_2 \rightarrow \infty$, for which $\Omega_2 \rightarrow 0$ and
${\cal U}_2/\Omega_2 \rightarrow 0$
by  (\ref{U-def}).

For discussion purposes, we  will take      
  ${\cal V} >0$.  
  This   is easily verified for the static
     asymptotically   RS  black string     
  \cite{black-string}, which is
 the only  known exact   solution for   an
  asymptotically   RS 
  black  object in 5-dimensional spacetime.  
 We will also see   in (\ref{tensions-def}) below that
   ${\cal V} >0$   if  and only if a gravitational tension  ${\cal T}_0$ is positive.

The coefficient of $\delta P$  
  defines  a thermodynamic volume  $V_{\rm eff}$ in a black hole first law 
  \cite{KNAdS, bh-enthalpy, Gibbons-enthalpy}.
 For a static asymptotically AdS black hole,
it was found in   \cite{bh-enthalpy} that  
  $V_{\rm eff}>0$ is   the    volume  \emph{removed} by  
  the  black hole (the        volume of pure AdS space 
minus the   volume outside   the black hole).
In  our first law,     $V_{\rm eff}=-{\cal V}<0$    
  suggests        that
 net  volume  is  \emph{added}   outside  the black hole (compared to  the case with no black hole).
 Added volume makes sense physically:  in RS2     the black hole  repels   the  positive-tension brane,
and  in  RS1 we  would  expect  a  version of 
 the    black hole  Archimedes effect   \cite{Archimedes-1, *Archimedes-2, Archimedes-3-tension}, where
 the    black hole     increases  the   size of the compact dimension (here  
 the    interbrane distance).
We also note that  $V_{\rm eff}<0$          
    occurs, with a natural interpretation
as an added volume,        in   AdS-Taub-NUT-AdS spacetime  \cite{negative-volume}.

In  RS1,     
there are three ways  
 the variation  $\delta L$  of the interbrane distance can be introduced into the first law, using
 (\ref{dL}).
In each case, the  coefficient of $\delta L$ defines a gravitational tension ${\cal T}$
that depends on which quantities are held fixed.
The three  gravitational tensions    we refer to below 
are  
 \begin{equation}
\label{tensions-def}
  {\cal T}_0 =    \frac{2P\,\cal V}{L} 
            \quad , \quad
    {\cal T}_1 = \frac{{\cal U}_1}{\ell}
        \quad , \quad
    {\cal T}_2 = \frac{{\cal U}_2}{\ell} \ .
\end{equation}
Using (\ref{dL}) in (\ref{first-law-v1}) to change variables from $\delta\ell$  to
 $\delta L$     gives 
 \begin{equation}
  \label{first-law-v2}
   \delta M = \frac{\kappa \,\delta A}{8\pi G_5}
 +    {\cal T}_0  \,\delta L
 + \sum_{i=1,2} \pm \Big({\cal U}_i -  {\cal T}_0\,\ell \Big) \frac{\delta\Omega_i}{\Omega_i}  \ .
\end{equation}
Here $\pm$ is   the   sign of    each brane tension  $\lambda_i$  and
 ${\cal T}_0$   is 
 a gravitational  tension   at
  fixed   values of   ($A$, $\Omega_1$, $\Omega_2$). 
Using   (\ref{dL})  in  (\ref{first-law-v1}) to  eliminate  $\delta \Omega_1$ or  $\delta \Omega_2$
gives  the first law as
 \begin{equation}
   \label{first-law-v3}
   \delta M = \frac{\kappa \,\delta A}{8\pi G_5}
 +    {\cal T}_1\,  \delta L
 +  \left(-{\cal V} +  \frac{{\cal T}_1 \, L}{2P} \right) \delta P
 + M \frac{\delta\Omega_2}{\Omega_2} 
\end{equation}
and
 \begin{equation}
    \label{first-law-v4}
   \delta M = \frac{\kappa \,\delta A}{8\pi G_5}
 +    {\cal T}_2\,  \delta L
 +  \left(-{\cal V} +  \frac{{\cal T}_2 \, L}{2P} \right) \delta P
 + M \frac{\delta\Omega_1}{\Omega_1}  \ .
\end{equation}
Each
 term  ${\cal T} \delta L$  
    in   (\ref{first-law-v2})$-$(\ref{first-law-v4})
is the work needed to  vary  the RS1 interbrane distance 
   (with different quantities   held fixed), 
  analogous to   the work terms in the first law
in   the case of a compact dimension
  without branes \cite{harmark-obers-def-tension, harmark-obers-first-law, kol-sorkin-piran}.
Our gravitational tensions  (\ref{tensions-def}) are   
   easily verified to be positive for  the      asymptotically   RS1   black string     
  \cite{black-string}. 
We would  also expect our gravitational tensions to    be positive 
  due the black hole's attraction to 
its   images in the covering space,
which has  been shown
   \cite{Archimedes-3-tension, tension-interpretation,positive-tension}
  in the case of   a compact dimension 
   without branes.

Since each version of     the first law
   reparametrizes   the  geometry,     
    in (\ref{first-law-v3}) and (\ref{first-law-v4}) 
  each thermodynamic volume     
$V_{\rm eff}= -{\cal V}+{\cal T}_i L/P$ 
differs  from $-\cal V$.
For $-{\cal V}<0$,    this indicates that 
positive gravitational tension ${\cal T}_i$
opposes the black hole Archimedes effect,
 and  the sign of   $V_{\rm eff}$ depends on  their     relative strengths.

Reparametrizing  the geometry
 also transforms the  brane terms in the first law,
but   with the interesting property      
 that the coefficients of $\delta\Omega_i/\Omega_i$   
  always add to $M$, in each version of the first law.
 A  brane term  
in (\ref{first-law-v1})  or (\ref{first-law-v2})  shifts  into  both
$V_{\rm eff}\, \delta P$ and       ${\cal T}_i \,\delta L$
in  (\ref{first-law-v3}) and (\ref{first-law-v4}),
and this shift   incorporates the brane's   orbifold   symmetry
into the gravitational tension,
since ${\cal T}_i$
   is due to the 
  black hole's   
  attraction to its orbifold mirror   images.

\section{Variational principle \label{vp-proof}}

The first law     we proved  in   section \ref{sec:first-law}  includes the variations of
the    AdS curvature length, cosmological constant, brane tensions, and RS brane warp factors.
This    motivates the   following
 variational principle,   which we   prove in this section.

  {\bf Variational principle for  asymptotically  RS black holes:} 
  {\it 
Instantaneously   static   initial data that 
extremizes the mass    $M$ is  initial data for
a static    black hole, for   variations
  that leave fixed the apparent horizon area $A$, 
  the AdS curvature length $\ell$, cosmological constant $\Lambda$, brane tensions  $\lambda_i$, 
and RS values (at spatial infinity) of the warp factors   $\Omega_i$ on each brane.}

\subsection{Main  proof \label{main step}}

Our proof of the variational principle proceeds in two steps.
Here we perform the main step, which  
  reduces the proof to   
  two auxiliary boundary value problems.  These boundary value problems are the topics of
  section \ref{BVPs}.
  Our setup is general: it applies to a black hole  localized on   a brane,   or    isolated    in the bulk (away from   either brane),  and also applies to the asymptotically RS black string \cite{black-string}.

Our  key assumptions  will be the following.  We   assume our
     initial data  $h_{ab}$   is instantaneously static.  We  also assume  
the variations  $\delta h_{ab}$          extremize the mass ($\delta M=0$), while
 holding fixed      the apparent horizon area $A$ and   the remaining quantities
   ($\ell$,   $\Lambda$,    $\lambda_i$, $\Omega_i$).      
 
 Our proof   closely follows the proof of the first law in     section \ref{sec:first-law}. 
 The initial steps    are the same, as we indicated after   the result (\ref{delta_H_sigma-1b}).
Thus, for  an instantaneously static
   geometry $h_{ab}$, we proceed exactly as in section \ref{sec:first-law} through (\ref{delta_H_sigma-1b}).
Now taking $\delta\Lambda=0$  in
  (\ref{delta_H_sigma-1b})     gives
\begin{equation}
\label{vp-delta_H_sigma}
\int_{\Sigma} d^{D-1}x \,{\cal P}^{ab}\delta h_{ab}
+
\sum_B I_B
=0 \ .
\end{equation}
In what follows, (\ref{vp-delta_H_sigma})
  will be our primary equation, 
where
\begin{equation}
\label{vp-Pab-initial}
 {\cal P}^{ab}   =   \frac{\sqrt{h}}{16\pi G_D}
  \left(
{\cal R}^{ab}+ h^{ab}D_c D^c  - D^a D^b
  \right)N \ .
\end{equation}
For   instantaneously static initial data, 
the   constraint ${\cal C}_0=0$  in (\ref{constraints}) simplifies to
  ${\cal R}=2\Lambda$
and $\delta {\cal C}_a$ vanishes identically. The  
 remaining  linearized constraint ($\delta{\cal C}_0=0$) 
simplifies to
\begin{equation}
\label{vp-linearized-constraint-1-red}
 \left({\cal R}^{ab} + h^{ab}D^c D_c -D^a D^b\right)
 \delta h_{ab}  = 0 \ .
\end{equation}

We now evaluate the  boundary  terms $I_B$ in (\ref{vp-delta_H_sigma}).
The boundaries $B$ are illustrated in Fig.\ \ref{initial-data-figure}.
 In section \ref{sec:first-law}, we evaluated     $I_{B_i}$   (at each brane) and 
$I_{B_\infty}$ (at spatial infinity), including the
variations of    quantities ($\ell$, $\Lambda$, $\lambda_i$, $\Omega_i$)   
  held  constant here  by  assumption. In this case,
(\ref{da}) and (\ref{text-I-infty-3})   reduce to
\begin{equation}
\label{surface-terms-values}
I_{B_i} = 0  \ , \quad
I_{B_\infty}= -\delta M    \ .
\end{equation}
Additionally, we have $\delta M=0$, by    our   assumption  
 of a mass extremum, so $I_{B_\infty}=0$.
 At the  apparent
 horizon $B_H$,
we use an  
 alternate form to that given  in
(\ref{surface-terms-2}), 
\begin{equation}
\label{surface-terms}
 I_{B_H} =  \int d^{D-2}x  \sqrt{\sigma} 
{\cal A}^{bcd}  \left[(D_bN)\delta h_{cd}
- ND_b\delta h_{cd} \right]  \ ,
\end{equation}
where 
\begin{equation}
16\pi G_D \, {\cal A}^{bcd}= n_a(h^{ac}h^{bd} -h^{ab}h^{cd}) \ .
\end{equation}
The boundary condition  on the lapse is $N=0$, whence
$D_aN = -f n_a$, where
$f^2 =  (D^bN)(D_bN)$.
Then     
  (\ref{surface-terms}) is
\begin{equation}
 I_{B_H}=
 \frac{1}{8\pi G_D} \int  d^{D-2}x \, f\,
\delta\sqrt{\sigma} \ ,
\end{equation}
using 
$\sigma_{ab}=h_{ab}- n_a n_b$
and 
$
 \delta \sqrt{\sigma}
=  \sqrt{\sigma}\sigma^{ab}\delta \sigma_{ab}/2
$.
For convenience, we  now  choose to set $I_{B_H}=0$
  using the following gauge transformation,
\begin{equation}
\label{horizon gauge}
\delta\sigma_{ab}  \rightarrow  \delta\sigma_{ab}   +   2{\cal D}_{(a}\xi_{b)} \ ,
\quad
\sigma^{ab}\delta \sigma_{ab}
   \rightarrow
 0  \ ,
\end{equation}
where ${\cal D}_a$ is the covariant derivative associated
 with $\sigma_{ab}$.
If we let $\xi_a={\cal D}_aF$, then 
$
\sigma^{ab}\delta \sigma_{ab}
   \rightarrow
 0
$
requires
\begin{equation}
\label{vp-throat-gauge}
-\sqrt{\sigma}\,{\cal D}^a{\cal D}_a F = \delta \sqrt{\sigma} \ .
\end{equation}
Note the apparent horizon is a closed
surface (this is most clearly seen in the  covering space, if the apparent horizon  intersects a brane).
A solution $F$ to (\ref{vp-throat-gauge}) on a closed surface is well known to exist
if  and only if the surface integral of the right-hand side of  (\ref{vp-throat-gauge})  vanishes.  This  integral 
  is   simply   $\delta A$, which indeed vanishes    since we   hold  $A$ constant.  Thus     
  a solution   $\xi_a$ exists to achieve (\ref{horizon gauge}), and we henceforth set $I_{B_H}=0$.
Since  $I_{B_i}=I_{B_\infty}=0$ and we      set $I_{B_H}=0$,
  our primary equation (\ref{vp-delta_H_sigma})    simplifies to
\begin{equation}
\label{vp-no-go-converse-1}
\int_{\Sigma} d^{D-1}x  \, {\cal P}^{ab}\delta h_{ab} =0
  \ .
\end{equation}

Our goal is   to conclude   that  the
initial  geometry $h_{ab}$
evolves to a static spacetime. The well known condition for this is 
 that ${\cal P}^{ab}=0$
on   $\Sigma$.  
We cannot, however, immediately conclude that  ${\cal P}^{ab}=0$  from (\ref{vp-no-go-converse-1}),
 because not all of the variations $\delta h_{ab}$
are arbitrary: the
linearized constraint  (\ref{vp-linearized-constraint-1-red})   removes one
degree of freedom, which  can be taken
  as       $h^{ab}\delta h_{ab}$  or as
  the variation $\delta h$ of the
 determinant $h$. These    are related by
$
h^{ab} \delta h_{ab}
=
 \delta h/h$.  
 As an identity, we may  decompose $\delta h_{ab}$ into a trace-free ($TF$)
part and a part proportional to $\delta h$:
\begin{equation}
\label{variations-decomp}
 \delta h_{ab} =
(\delta h_{ab})^{TF}
  +
\frac{1}{D-1}\left(\frac{\delta h}{h}\right)h_{ab} \ .
\end{equation}
Using   (\ref{variations-decomp}),
our primary equation (\ref{vp-no-go-converse-1})  then  becomes
\begin{equation}
\label{vp-no-go-converse-2}
 \int_{\Sigma} d^{D-1}x
 \left[ ({\cal P}^{ab})^{TF}(\delta h_{ab})^{TF}
  +
 \frac{{\cal P}\, \delta h}{(D-1)h}
  \right]
  =0 \ ,
\end{equation}
where   
\begin{eqnarray}
\label{P-decomp}
  {\cal P}_{ab} &=&
( {\cal P}_{ab})^{TF}
  +
\frac{\cal P}{D-1} h_{ab}  \ ,
\\
{\cal P} &=& h_{ab}{\cal P}^{ab} \ .
\end{eqnarray} 
The arbitrary  variations are $(\delta h_{ab})^{TF}$,  
subject to  
  smoothness at the apparent horizon,
  $(\delta h_{ab})^{TF} \rightarrow 0$ at $B_\infty$,
  and       boundary conditions at the branes that we will specify 
   in the next section.
 As a completeness check,  the  arbitrary  variations  $(\delta h_{ab})^{TF}$
   alone should 
determine the dependent quantity  $\delta h$, which
we   verify below by showing the
linearized constraint   (\ref{vp-linearized-constraint-1-red}) 
is a well posed boundary value problem  for $\delta h$.

Our proof   then  reduces to 
showing     
     ${\cal P}=0$, which  allows 
us to conclude  
from (\ref{vp-no-go-converse-2}) that
 $({\cal P}^{ab})^{TF}=0$, since
  $(\delta h_{ab})^{TF}$
are   arbitrary variations.
It   then follows  from (\ref{P-decomp}) that
  ${\cal P}^{ab}=0$, which is the desired result.
The statement  ${\cal P}=0$ 
is a  boundary value problem for $N$ that we demonstrate   
 is solvable in the following section,   which
completes our proof of the variational principle.

\subsection{Auxiliary boundary value problems \label{BVPs}}

The  boundary value problem for  $N$     is 
   \begin{subequations} \label{N-BVP}
\begin{eqnarray}
\label{vp-N-eom}
 D^a D_a N  -    \frac{(D-1)}{\ell^2}  N
 &=& 0 \ ,
 \\*
\label{vp-N-bc-brane}
\left(n^a D_a   N   -      \frac{\varepsilon}{\ell}\, N \right)\Big|_{B_i}
  &=&   0  \ ,
  \\*
  \label{vp-N-bc-throat}
    N \, \Big|_{B_H} &=&   0  \ ,
    \\*
    \label{vp-N-bc-infty}
    N\, \Big|_{B_\infty} &\rightarrow& \Omega  \ .
\end{eqnarray}
   \end{subequations}
Here,   $\Omega$ is the warp factor of the asymptotic RS solution (\ref{g_RS}) and
$\varepsilon=\pm 1$ is the sign of each brane tension.
The result (\ref{vp-N-eom})  follows from  setting
  ${\cal P}= h_{ab}{\cal P}^{ab}=0$,
  using
(\ref{vp-Pab-initial}) and
    ${\cal R}=2\Lambda$.
The boundary conditions (\ref{vp-N-bc-throat}) and   (\ref{vp-N-bc-infty})
are straightforward.
Our main concern is the brane boundary condition   (\ref{vp-N-bc-brane}), which
 results from  using
\begin{equation}
2K_{ab} = n^c \partial_c \gamma_{ab} + \gamma_{ac} \partial_b n^c +  \gamma_{bc} \partial_a n^c \ .
\end{equation}
Now    $\gamma_{tt}=-N^2$ and $\gamma_{ta}=0$  gives
$
2K_{tt} = n^c \partial_c (-N^2) 
$,
and the Israel condition $K_{tt} = (\varepsilon/\ell)\gamma_{tt}$ then   gives
 (\ref{vp-N-bc-brane}).

 As shown in  \cite{inverse-positivity}, the following approach
  can    put a Robin boundary condition (\ref{vp-N-bc-brane})   in a standard   form while keeping
    its associated elliptic equation  (\ref{vp-N-eom})  in a   divergence form.
Let $w_a$ be any vector field and  define   ${\cal W}_aN = (D_a  - w_a)N$.
Then  (\ref{vp-N-eom}) and  (\ref{vp-N-bc-brane})  
become
  \begin{equation}
  \label{vp-N-eom-2}
 D^a {\cal W}_a  N 
 + w^a D_a N
+  \left[    D_a w^a -  \frac{(D-1)}{\ell^2}\right] N
=0 
\end{equation}
and
\begin{equation}
\label{vp-N-bc-brane-2}
\left[n^a {\cal W}_a N+ \left(n^a w_a  -  \frac{\varepsilon}{\ell}\right)N \right]\Big|_{B_i}
=   0 \ .  
\end{equation}
As in   \cite{inverse-positivity}, we now
 choose $w_a$ so      $(n^aw_a - \varepsilon/\ell) \ge 0$ at     $B_i$,
which is the usual prerequisite for applying an existence theorem  to a    boundary 
   value problem of the form
    (\ref{vp-N-eom-2})\textendash(\ref{vp-N-bc-brane-2}).
For example, we choose $w_a = - \widetilde n_a/\ell$,
where $\widetilde n_a$  is 
any  vector field,
  pointing from   
   $B_1$ to   $B_2$,
    that   interpolates
    from the inward unit normal  ($-n_a$) of  $B_1$
to the
outward unit normal $n_a$  of $B_2$. 
Then $n^a\widetilde n_a = -\varepsilon$ at each brane $B_i$,
and
$w_a = - \widetilde n_a/\ell$     gives $(n^aw_a - \varepsilon/\ell) = 0$ in  (\ref{vp-N-bc-brane-2}).
With the brane  boundary conditions in  standard form,  and   the remaining
 standard (Dirichlet) boundary conditions, 
  (\ref{vp-N-bc-throat}) and   (\ref{vp-N-bc-infty}),
 we   then  readily infer that the  boundary value problem
(\ref{N-BVP}) for $N$ is solvable.

We now turn to the boundary value problem for
  $\delta h$, 
which  we will state in terms of a scalar quantity $(\delta h/h)$,
   \begin{subequations} \label{dh BVP}
\begin{eqnarray}
\label{vp-dh-eom}
 D^a D_a (\delta h/h)   -  \frac{(D-1)}{\ell^2}  \, (\delta h/h)
 &=& f_{\Sigma} \ ,
 \\*
\label{vp-dh-bc-brane}
\left[n^a D_a   (\delta h/h)   -    \frac{\varepsilon}{\ell} \, (\delta h/h)
\right]\Big|_{B_i}
  &=&   f_{i} \ ,
  \\*
\label{vp-dh-bc-throat}
n^a D_a (\delta h/h)   \Big|_{B_H} &=&   f_H \ ,
    \\*
 \label{vp-dh-bc-infty}
    (\delta h/h)\ \Big|_{B_\infty} &\rightarrow& 0 \ .
\end{eqnarray}
   \end{subequations}
   As above,  $\varepsilon =\pm 1$ is the sign of each brane tension.
The  result   (\ref{vp-dh-eom})   follows from
 substituting  (\ref{variations-decomp})
 into the  linearized
constraint 
(\ref{vp-linearized-constraint-1-red})
 with ${\cal R}=2\Lambda$.
We will give the source terms and
  derive the boundary conditions below.

The key point  is that
(\ref{dh BVP}) is a well posed 
boundary value problem. The  
 elliptic equation 
  (\ref{vp-dh-eom})   and the   boundary conditions 
   (\ref{vp-dh-bc-brane})  
 are  similar in  form to 
   (\ref{vp-N-eom}) and  (\ref{vp-N-bc-brane})  in the  previous  boundary value 
   problem (\ref{N-BVP}). The remaining boundary conditions,
 (\ref{vp-dh-bc-throat}) and (\ref{vp-dh-bc-infty}), are   well known types:
 Neumann and  Dirichlet, respectively.

In the remainder of this section, we
provide the details of the source terms and
  the boundary conditions in (\ref{dh BVP}).
The source terms in (\ref{dh BVP}) are  
\begin{eqnarray*}
   f_{\Sigma} &=&
  \frac{D-1}{D-2} \left[D^a D^b(\delta h_{ab})^{TF}
   -
  {\cal R}^{ab} (\delta h_{ab})^{TF} 
  \right]\ ,
\\*
 -f_{i} &=&  \frac{D-1}{D-2} 
 \left[
\sigma^{ab} n^c D_c (\delta
h_{ab})^{TF} +  \frac{\varepsilon D}{\ell}n^a n^b (\delta h_{ab})^{TF}
 \right]
 \ ,
\\*
  f_H &=&
  \frac{1}{D-2}
  \left[
2k^{ab} (\delta h_{ab})^{TF} - \sigma^{ab} n^c D_c (\delta h_{ab})^{TF}
\right] \ .
\end{eqnarray*}
 The boundary conditions on $\delta h$  are given
  by
varying those  on $h_{ab}$, which
at the apparent horizon and the branes 
 involve the extrinsic curvature $k_{ab}=\sigma_a{}^c D_c n_b$,
\begin{equation}
\label{k-boundary-conditions}
k \Big|_{B_H} =   0
\ , \quad
k_{ab} \Big|_{B_i} = \frac{\varepsilon}{\ell} \, \sigma_{ab} \ .
\end{equation}
By varying these, we obtain
\begin{equation}
\label{unconstrained-var}
\delta k \Big|_{B_H} =   0
\ , \quad
\delta k \Big|_{B_i} =   0
\ , \quad
\delta k_{ab} \Big|_{B_i} = \frac{\varepsilon}{\ell} \, \delta \sigma_{ab} \ .
\end{equation}
 To evaluate these,
we use the   general results
   \begin{subequations}
\begin{eqnarray}
 \label{unconstrained-var-dk_ab}
2\delta k_{ab}  &=&
 \left(n^c n^d\delta h_{cd}\right)k_{ab}
 -
 \sigma_a{}^c \sigma_b{}^d  n^f
J_{cdf} \ ,
\\
\label{unconstrained-var-dk}
-2\delta k  &=&  2k^{ab} \delta \sigma_{ab}  -  k\,n^a n^b \delta h_{ab}
+   \sigma^{ab} n^c J_{abc} \ ,
\\
J_{abc} &=&  D_{a}\delta h_{bc} + D_{b}\delta h_{ac} - D_c\delta h_{ab} \ .
\end{eqnarray}
   \end{subequations}
The boundary conditions at the branes  (\ref{vp-dh-bc-brane})  
and the apparent horizon   (\ref{vp-dh-bc-throat}) 
result  from
evaluating  $\delta k=0$ using
 (\ref{variations-decomp}),
 (\ref{k-boundary-conditions}),
 and
 (\ref{unconstrained-var-dk}).
The last relation in
 (\ref{unconstrained-var})
expresses
brane  boundary conditions for $(\delta h_{ab})^{TF}$, since
it
reduces
to a form   independent of $\delta h$
after
substituting
(\ref{RS-K}),
(\ref{variations-decomp}),
  (\ref{vp-dh-bc-brane}), and (\ref{unconstrained-var-dk_ab}).

\subsection{An application of the variational principle \label{black string application}}

Here we    
  demonstrate  the utility of the variational principle, 
  by applying it   to a trial  solution
and reproducing
   the static asymptotically   RS    black string        \cite{black-string},
which is  the only  known exact     solution for   an
  asymptotically   RS  
  black  object  in 5-dimensional spacetime. 
We first  specify  a  trial  geometry for an initially static a black string. After evaluating
the apparent horizon area $A$ and mass $M$, we   
then   apply the variational principle.

A   black string  
is a set of lower dimensional black holes   stacked in an extra dimension $Z$,
which is how we    will   construct the trial geometry.  We take
\begin{equation}
\label{ansatz}
ds^2 = 
\Omega(Z)^2
\left[
\Psi^4({\bf x},Z) \, d{\bf x}^2+ dZ^2
\right] \ ,
\end{equation}
where $\Omega=\ell/Z$ and the branes are  the surfaces
  $Z=Z_1$ and $Z=Z_2$.
We  will  take
\begin{equation}
\label{Psi-1}
\Psi = 1 + \frac{\rho_0}{2}\left(
\frac{1}{|{\bf x}+{\bf x}_0|} +  \frac{1}{|{\bf x}-{\bf x}_0|}
\right) \ .
 \end{equation}
Here   $\rho_0 >0 $  is a constant and
 ${\bf x}_0= (0,0, d)$, using
  Cartesian coordinates ${\bf x} =(x, y, z)$ 
with
    origin at ${\bf x}=0$.  
    We  choose  a function $d(Z)$ as follows.
The constraint, ${\cal R}= -12/\ell^2$, after linearizing
 in $d$ and its derivatives, has the solution $d(Z) =d_0 + c_0 Z^4$, where
$c_0$ and $d_0$ are constants.  
For the case $c_0=0$,     (\ref{Psi-1})   is an exact solution
and  the   RS1 limit ($\Omega_2\rightarrow 0$) is easily taken.
We will work in RS2 and take  $c_0$  as a small nonzero parameter.

    On each slice $Z$=constant, we  now transform to 
  spherical coordinates centered at ${\bf x}=0$, with $z=\rho \cos\theta$,
and
 expand $\Psi$  in    Legendre polynomials $P_{k}(\cos\theta)$  for $\rho > d$,
 \begin{equation}
  \label{Psi-2}
 \Psi =1 + \frac{\rho_0}{\rho} +  
\frac{\rho_0}{\rho}  \sum_{j=1}^\infty \left(\frac{d}{\rho}\right)^{2j} P_{2j}(\cos\theta) \ .
\end{equation} 
 On each slice $Z$=constant,  we  will  take   $\rho_0 \gg d$, which
  describes a
 3-dimensional   black hole    \cite{Brill-Lindquist}, and   $d$  parametrizes  
   the    2-dimensional apparent horizon's distortion    from the sphere $\rho=\rho_0$.
The  full   3-dimensional apparent horizon   
       (the union of   the 2-dimensional 
 apparent horizons) therefore
   describes a  distorted black string.

 On each slice $Z$=constant,  as in  \cite{Brill-Lindquist}, 
the  surface $\rho(\theta)$  of the 2-dimensional    apparent horizon    can be found as the sum of  Legendre polynomials  
   that 
  minimizes the area $A_2$,
 \begin{equation}
A_2 = 2\pi    
\int_0^{\pi}  d\theta\, \Psi^4 \rho \sqrt{\rho^2 + \left( \frac{d\rho}{d\theta}\right)^2} \ .
 \end{equation}
To lowest order in  $(d/\rho_0) \ll 1$,   we   find the  apparent horizon
on each slice $Z$=constant is
  \begin{equation}
  \label{bs-AH}
 \rho(\theta) = \rho_0
 \left[ 1  +  \frac{5}{7}\left(\frac{d}{\rho_0} \right)^2 P_2(\cos\theta) \right] \ .
 \end{equation}
 This agrees  with the numerical results of \cite{Brill-Lindquist},
and gives, to lowest order  in $(d/\rho_0$),
\begin{equation}
\label{bs-area-2D}
A_2(Z)  =  
64\pi \rho_0^2 
\left[
1 - \frac{5 }{7}\left(\frac{d}{\rho_0}\right)^4
\right] \ .
 \end{equation}
 Integrating in  $Z$ gives
the 3-dimensional
 area of the   full  apparent horizon in the   black string geometry,     
\begin{equation}
 \label{bs-area}
A =      \int_{Z_1}^{Z_2} dZ\, \Omega^3   
A_2  
=
32\pi  w 
\left(
\rho_0^2  - \frac{5Q}{7\rho_0^2}  
\right) \ ,
 \end{equation}
where, with   $\Omega_i$   the warp factor  at each brane,
  \begin{equation}
 w = \ell \left(\Omega_1^2 - \Omega_2^2\right)
 \ , 
\quad
Q =  d_0^4 
+
\frac{8 \, c_0 \, d_0^3 \, \ell^4}{\Omega_1\Omega_2(\Omega_1+\Omega_2)} \ .
 \end{equation}
We defined the 
 mass $M$   in (\ref{mass-geometric}), which for
(\ref{Psi-2})    gives
  \begin{equation}
  \label{bs-mass}
 G_5 M =w  \rho_0 \ .
 \end{equation}
To lowest order in $Q$, combining  (\ref{bs-area})\textendash(\ref{bs-mass})  gives
\begin{equation}
\label{bs mass function}
\left(\frac{G_5 M}{w}\right)^2 =  \left(\frac{A}{32\pi w } \right)+ \frac{5}{7} \left(\frac{32\pi w }{A} \right)  Q \ .
 \end{equation}
We   now     apply our variational principle: we extremize 
  $M$  in (\ref{bs mass function}) at fixed  $A$, $\ell$, and $\Omega_i$. 
  This yields the conditions $c_0=0$ and $d_0=0$, which   we conclude   
  describes  a static black string. 
  We can verify this    directly,
 since  $d=0$ in  (\ref{Psi-2}) gives
   $\Psi = 1+\rho_0/\rho$ and the apparent horizon 
 (\ref{bs-AH})
 is located at $\rho=\rho_0$.  This is indeed the initial geometry
of  the static black string  \cite{black-string}
 in isotropic coordinates.

 We can also deduce that 
 the   evolution of  the distorted black string, with $d \neq 0$, will not be static, since    
 each slice $Z$=constant is   the initial geometry for an attracting
  two-body problem  \cite{Brill-Lindquist}, and
 the 2-dimensional apparent horizon   considered above    
  describes a black hole formed by, and surrounding,
  two closely separated smaller black holes (with 
a small minimal surface  surrounding each   point  ${\bf x}=\pm{\bf x}_0$).
From the perspective in each slice $Z$=constant,
as in    \cite{Brill-Lindquist}, the
 two small interior black holes 
  will coalesce   as
  the initial data evolves, due to   
  mutual gravitational attraction.  This
     results in a time-dependent geometry on each slice $Z$=constant,
 and   results in    a time-dependent black string 
      geometry in the bulk perspective.

\section{Conclusion \label{sec:conclusion}}

We have derived the first law for a static asymptotically RS black hole,   whose 
mass  $M$   is defined in (\ref{mass-geometric}) and (\ref{mass}).
Four  versions of this law are   
  given in  (\ref{first-law-v1})\textendash(\ref{first-law-v4})  for different choices of variations.
 In both RS1 and RS2, the general 
 first law contains brane terms and a thermodynamic volume.
In           RS1,   we can define both a    thermodynamic volume and a 
  gravitational tension,
due to the  presence  of  both 
a cosmological constant  and a compact interbrane distance.
This differs from   the first law in previously studied spacetimes
(with  either a cosmological constant
or a compact dimension), 
where the analogs of our thermodynamic volume and gravitational tension
 are isolated from each other, appearing in    the   separate  first laws   of      separate spacetimes.

  The  variational principle   we developed in this paper states that
  for an asymptotically RS black hole
  initially at rest,   initial data that extremizes the mass yields a static black hole,
   for variations at fixed values of the apparent horizon area and the 
   remaining physical variables in the first law ($L$, $\Omega_i$, $\ell$, $\Lambda$,   $\lambda_i$).
 It would be interesting to investigate the consequences of holding fewer variables  fixed.
An example of this  in four-dimensional spacetime
 is Hawking's proof     \cite{hawking}    that the
  static (Schwarzschild) black hole is an extremum of mass   
at fixed apparent horizon area but   arbitrary angular momentum. 
 
Our example application of the   variational principle   
 to a trial solution serves as
a prelude to the approach we will take in a sequel paper \cite{paper-3-extrema}.
In \cite{paper-3-extrema}, we
will  
 conclude that solutions exist for small static black holes in RS2, 
 both on and off the brane, as special members of 
 a general family of initially static black holes.
This family of black hole initial data will also indicate   
  that a   small black hole  on an orbifold-symmetric brane in RS2   is stable against leaving the brane, 
which generalizes 
  to other    models with an orbifold-symmetric brane.
 If we inhabit such a brane, then small    black holes, if  produced in   high energy collider experiments on the brane,    could  be studied   directly  (instead of leaving behind a signature of missing energy), which is an important result for future  experiments at the LHC.

\begin{acknowledgments}
   We thank D.\ Kastor and   J.\ Traschen for useful discussions at KITP.
   This research was supported in part by the National Science Foundation
   under Grant No.\ NSF PHY11-25915.
\end{acknowledgments}

\appendix
\section{Mass counterterm $M_{ct}$}
\label{appendix-M-ct}

In
this appendix, we derive the mass counterterm $M_{ct}$ 
from (\ref{M-ct-formula}) and (\ref{S-ct-hybrid}), and thereby
prove the mass formula (\ref{mass-geometric}). 
We begin with  the variation  of (\ref{S-ct-hybrid}),
\begin{equation}
\label{d S_ct}
\delta  S_{ct}
  =  - \int
d^4x
\frac{\sqrt{-\gamma} }{ 8\sqrt{2}\pi G_5}
\left(\sqrt{\hat{\cal R}}
\gamma^{ab} \delta\gamma_{ab}
   +   \frac{\delta \hat{\cal R}}{\sqrt{\hat{\cal R}}}    \right)
 \ .
\end{equation}
The standard   variation of the Ricci scalar
is  
\begin{equation}
  \delta{\hat{\cal R}} = - {\hat{\cal R}}^{ab}\delta\hat\sigma_{ab}    +  \hat{d}_a v^a
\end{equation}
where  $v^a = 2 \hat\sigma^{b[a}\hat{d}^{c]}\delta\hat\sigma_{bc}$
and $\hat{d}_a$ is the covariant derivative associated with
$\hat\sigma_{ab}$. This gives
\begin{equation}
\label{A3}
\delta S_{ct} = \int d^4x\, {\cal S}^{ab} \delta\gamma_{ab} - (8\sqrt{2}\pi G_5)I \ ,
\end{equation}
where ${\cal S}^{ab}$ is given below and,
  with $J=\sqrt{-\gamma_{tt}\gamma_{ZZ}/\hat{\cal R}}$,
\begin{equation}
  I
=
  \int dt\,dZ\,
\int d^2x \, \sqrt{\hat\sigma}\,
\left[
\hat {d}_a (J v^a) -v^a \hat {d}_a J
\right]
 \ .
\end{equation}
We conclude  that $I=0$, as follows.  The first term  is a total divergence, but the 2-sphere has  no boundary. Also,
$\hat {d}_a J=0$
since we take  $J$  independent of
the angular coordinates.  Then   (\ref{A3}) gives
\begin{equation}
 {\cal S}^{ab}  = \frac{\delta S_{ct}}{\delta\gamma_{ab}} \ .
\end{equation}
In (\ref{d S_ct}) we now use
\begin{equation}
  \gamma^{ab}\delta\gamma_{ab}  =   \gamma^{tt}\delta\gamma_{tt}
+ \hat\sigma^{ab}\delta\hat\sigma_{ab}
+ \sigma^{ZZ}\delta\sigma_{ZZ}
\end{equation}
which gives
\begin{subequations}
\begin{eqnarray}
\label{fderiv-S-ct-1}
 \frac{\delta S_{ct}}{\delta\gamma_{tt}} &=& 
  - \, \frac{1}{8\sqrt{2}\pi G_5}\, \sqrt{-\gamma\hat{\cal R}}
    \, \gamma^{tt} \ , \
\\
\label{fderiv-S-ct-2}  \frac{\delta S_{ct}}{\delta\hat\sigma_{ab}} 
&=& 
  - \, \frac{1}{8\sqrt{2}\pi G_5}\,
   \sqrt{-\gamma \hat{\cal R}} \left(
\hat\sigma^{ab}   -
\frac{\hat{\cal R}^{ab}}{\hat{\cal R}}
\right) \ , \qquad  
\\
\label{fderiv-S-ct-3}  \frac{\delta S_{ct}}{\delta\sigma_{ZZ}} &=& 
  - \, \frac{1}{8\sqrt{2}\pi G_5}\,
   \sqrt{-\gamma \hat{\cal R}} \, \sigma^{ZZ} \ .
\end{eqnarray}
\end{subequations}
The mass counterterm from (\ref{M-ct-formula}) and
 (\ref{fderiv-S-ct-1}) is then 
\begin{equation}
\label{M-ct}
  M_{ct}
 =
   \frac{\sqrt{2}}{8\pi G_5}
 \int d^3 x \,  \sqrt{-\gamma \hat{\cal R}} \ .
\end{equation}
Combining this with  (\ref{M-bare-formula}) now gives  
the mass formula
 (\ref{mass-geometric}).

\section{Boundary term $I_B$  at infinity}

\label{appendix-first-law}

Here we evaluate the term  $I_{B_\infty}$ 
given by  (\ref{surface-terms-2}) at the
boundary    $\rho \rightarrow \infty$.
Throughout this appendix,  $\simeq$ denotes evaluating  at leading order  and  neglecting terms of higher order  in  $1/\rho$.
We will relate  $I_{B_\infty}$ to the mass variation $\delta M$ and additional terms.
The mass is a sum of two terms, $M=\widetilde M+M_{ct}$,
whose individual variations  are
\begin{equation}
  \label{dM-tilde}
 \delta \widetilde M
 = -\, \frac{1}{8\pi G_5}\int_{B_\infty} d^{3}x\,\left[N
\delta\left(\sqrt{\sigma} \,k\right)  + \sqrt{\sigma}\, k\,    \delta
N \right]
\end{equation}
and
\begin{equation}
 \label{dM-ct-3}
 \delta M_{ct} =
 \int_{B_\infty} d^3 x \,  \left[\frac{\delta M_{ct}}{\delta\sigma_{ab}} \, \delta\sigma_{ab}
  +  \frac{ \sqrt{2\sigma  \hat{\cal R}}}{8\pi G_5}
  \, \delta N \right]  \ .
\end{equation}
Note
$\delta M_{ct}/\delta \sigma_{ab} =
-\delta S_{ct}/\delta \sigma_{ab}$
 since 
 $\int dt\,M_{ct}=-S_{ct}$
by (\ref{M-ct}) and
(\ref{S-ct-hybrid}).
From
(\ref{fderiv-S-ct-2}) and (\ref{fderiv-S-ct-3}), we find
 \begin{equation}
 \label{dM-ct-X}
  \frac{\delta M_{ct}}{\delta\sigma_{ab}} -
   \frac{N}{2} \sqrt{\sigma} s^{ab} =  X^{ab}  
\end{equation}
at large $\rho$,
where the quantities  $X^{ab}$ are given below.
Using (\ref{dM-tilde})\textendash(\ref{dM-ct-X}), we rewrite the boundary term
(\ref{surface-terms-2})  as
\begin{eqnarray}
\nonumber I_{B_\infty} &=&  -  \delta M
  +  \int_{B_\infty} d^{3}x \, X^{ab}\delta\sigma_{ab}
  \\
& &
  -  \frac{1}{8\pi G_5} \int_{B_\infty} d^{3}x \, \sqrt{\sigma}
 \left(k -\sqrt{2\hat{\cal R}}\right)\delta N \ . \qquad
\label{I-infty-2}
\end{eqnarray}
For the metric asymptotics (\ref{ansatz-1}), the quantities $X^{ab}$ are
\begin{subequations}
\begin{eqnarray}
 16\pi G_5\, X^{\chi\chi} &\simeq& \frac{\Omega \sin\chi}{\rho^2}
 \left(a_t+a_\rho+a_Z\right) \ ,
\\
 16\pi G_5\,  X^{\phi\phi} &\simeq& \frac{\Omega }{\rho^2\sin\chi}
\left(a_t+a_\rho+a_Z\right) \ ,
\\
\label{X_ab}
 16\pi G_5\,  X^{ZZ} &\simeq&  \Omega\sin\chi
\left( a_t +2a_\rho\right)    \ .
\end{eqnarray}
\end{subequations}
Here $\chi$ is the polar angle on the 2-sphere with radius $\rho$.
We now proceed to evaluate   (\ref{I-infty-2}).
We begin   with three convenient variables ($\ell$, $Z_1$, $Z_2$) and then   express
results in terms of   three  physical variables ($\ell$, $\Omega_1$, $\Omega_2$).
We first consider the variation  $\delta\ell$ at fixed ($Z_1$, $Z_2$).
At large $\rho$, we have $\delta g_{ab} \simeq 2(\delta\ell/\ell) g_{ab}$.
Then
\begin{equation}
  \int_{B_\infty} d^{3}x\, X^{ab}\delta\sigma_{ab} =
 {\cal F}_\infty \delta \ell
\end{equation}
where  
\begin{equation}
\label{F-infty-appendix}
{\cal F}_\infty =
\frac{1}{2G_5\ell}\int_{Z_1}^{Z_2} dZ\, \Omega^3\left( a_t  + 2a_\rho \right) \ .
\end{equation}
This is entirely due to $\delta \sigma_{ZZ}$ since the  integral contributions from
  $\delta\sigma_{\phi\phi}$ and   $\delta\sigma_{\chi\chi}$
vanish by the identity
(\ref{identity1}).  This identity can also be used to rewrite ${\cal F}_\infty$ in the form given in (\ref{F-infty}).
The
last line  in (\ref{I-infty-2})
 yields
  $M  \delta \ell/\ell$. Hence
  (\ref{I-infty-2}) yields, at  fixed ($Z_1, Z_2)$,
 \begin{equation}
 \label{I-fixed Z}
\left(I_{B_\infty}\right)_{Z_1 Z_2} = -  \delta M +
  \left({\cal F}_\infty +\frac{M}{\ell}\right) \delta \ell
  \ .
\end{equation}
We now consider   variations ($\delta Z_1$, $\delta Z_2$) at fixed $\ell$.
We   then perform a  coordinate transformation
 \begin{equation}
Z \rightarrow \widetilde Z = (1-\epsilon)Z - \zeta
\end{equation}
 such that the branes   again reside at $Z=Z_i$.  The required transformation is
 \begin{equation}
 \epsilon = \frac{\delta Z_2-\delta Z_1}{Z_2-Z_1}
\quad , \quad
\zeta = \frac{Z_2\,\delta Z_1- Z_1\,\delta Z_2}{Z_2-Z_1} \ .
\end{equation}
At large $\rho$, the resulting metric perturbation  is
 \begin{equation}
 \delta g_{ab}
\simeq
2\Omega^2\left[
 \epsilon\left(\delta^Z_a \delta^Z_b -   \eta_{ab}\right)
-\frac{\zeta}{\ell} \,\Omega\eta_{ab}\right]
  \ ,
\end{equation}
 where $\eta_{ab}$
is the 5-dimensional Minkowski metric.
Then (\ref{I-infty-2}) becomes, at fixed $\ell$,
 \begin{equation}
 \label{I-fixed-ell}
 \left(I_{B_\infty}\right)_\ell = -  \delta M
  - \epsilon M - \zeta\,{\cal I}
  \ ,
\end{equation}
where the integral $\cal I$   is
 \begin{equation}
{\cal I} =
\frac{3}{2G_5\ell}
\int_{Z_1}^{Z_2} dZ\, \Omega^4 \left(a_t + 2a_\rho + a_Z\right)  
  \ .
\end{equation}
For the case when all three quantities 
($\ell$, $Z_1$, $Z_2$) are varied,   we combine   (\ref{I-fixed Z})  and (\ref{I-fixed-ell}) to obtain
\begin{equation}
\label{I-infty-3}
 I_{B_\infty}  = -  \delta M +
  \left({\cal F}_\infty +\frac{M}{\ell}\right) \delta \ell
 - \epsilon M - \zeta\,{\cal I}  \ .
\end{equation}
We  can evaluate the integral $\cal I$  using 
(\ref{a_t}) and the identity
(\ref{int-field}).  We then express the result    
  in terms of  $M$ and $\cal Q$ using
 (\ref{M and mu0}) and
 (\ref{a-M-Q}).  This gives
  \begin{equation}
  \label{integral I}
 {\cal I} \ell = 
M\left( \frac{\Omega_1{}^3 -\Omega_1{}^3 }{\Omega_1{}^2-\Omega_2{}^2}\right)
-
\frac{3 {\cal Q}}{2G_5}
\left(
 \frac{\Omega_1{}^2\Omega_2{}^2}{\Omega_1{}^{2}-\Omega_2{}^{2}} 
\right) \ .
\end{equation}
We also express  $\epsilon$ and $\zeta$   in terms of three physical    variables ($\ell$, $\Omega_1$, $\Omega_2$)  using
\begin{equation}
\label{dZ}
\delta Z_i = \frac{1}{\Omega_i} 
\left( \delta\ell
-
\frac{\ell}{\Omega_i} \, \delta\Omega_i
\right) \ .
\end{equation}
Using (\ref{integral I}) and (\ref{dZ}) in  (\ref{I-infty-3})   then yields the result  for
$ I_{B_\infty}$   given in 
(\ref{text-I-infty-3}). 

\section{Gauge invariance}
\label{gauge invariance}
  
It is important to confirm that our quantities   
($M$, $\cal Q$, $\cal V$, $L$, ${\cal T}_0$, ${\cal T}_i$, ${\cal U}_i$)  are   gauge invariant at infinity.
As one   can verify, these quantities are    invariant under  the following   metric  transformation    
that leaves the branes fixed,
\begin{equation}
a_\nu \rightarrow a_\nu - \frac{\Omega^\prime}{\Omega}w  - \delta^Z_\nu w^\prime 
\ \;  , \ \; 
w(Z_1)=w(Z_2)=0
\ ,
\end{equation}
with $^\prime=d/dZ$.  This is generated
by   the coordinate transformation
$x^a  \rightarrow   x^a + \varepsilon^a$,
where
to leading order in $1/\rho$,
\begin{equation}
\varepsilon^Z = \frac{w}{\rho}  
\quad ,   \ \quad 
\varepsilon^\rho =  \frac{W}{\rho^2}  
\quad ,   \ \quad 
 w=W^\prime \ .
\end{equation}
In particular, we  consider the quantity ${\cal F}$, and write
\begin{equation}
 {\cal F} = {\cal F}_\Sigma + {\cal F}_\infty
\end{equation}
where
$ {\cal F}_\Sigma$ is 
  the sum of bulk  and brane terms in
 (\ref{F-def}),
 \begin{equation}
{\cal F}_\Sigma \,\delta\ell  \  \equiv \
 \frac{\delta\Lambda}{8\pi G_5}\int_{\Sigma} d^{4}x\, N
\sqrt{h}
 \ + \ I_{B_1} \ +\  I_{B_2} \ .
\end{equation} 
We note that  $\cal F$ is gauge invariant, 
but neither ${\cal F}_\Sigma$   nor ${\cal F}_\infty$  is separately 
     invariant, since
  they transform as 
  \begin{equation}
{\cal F}_\Sigma \rightarrow {\cal F}_\Sigma - \varphi
\quad , \quad
{\cal F}_\infty \rightarrow {\cal F}_\infty +   \varphi \ ,
\end{equation}
where
\begin{equation}
 \varphi  =
\frac{3}{2G_5\ell^2}\int_{Z_1}^{Z_2} dZ \,\Omega^4 w \ .
\end{equation}

\bibliography{first-law-vp}

\end{document}